\documentstyle[12pt]{article}
\input{epsf}
\newcommand{\beq}{\begin{equation}}
\newcommand{\eeq}{\end{equation}}
\newcommand{\be}{\begin{eqnarray}}
\newcommand{\ee}{\end{eqnarray}}
\begin{document}
\rightline{RUB-TPII-15/98}
\rightline{hep-ph/9809487}
\vspace{.3cm}
\begin{center}
\begin{large}
{\bf Flavor asymmetry of polarized and unpolarized
sea quark distributions in the large--$N_c$ limit$^\dagger$} \\[1cm]
\end{large}
\vspace{1.4cm}
{\bf B.\ Dressler}$^{\rm a}$, {\bf K.\ Goeke}$^{\rm a}$, 
{\bf P.V.\ Pobylitsa}$^{\rm a, b}$, {\bf M.V.\ Polyakov}$^{\rm a, b}$,
{\bf T. Watabe}$^{\rm a, c}$, {\bf and C. Weiss}$^{\rm a}$  
\\[1.cm]
$^{a}${\em Institut f\"ur Theoretische Physik II,
Ruhr--Universit\"at Bochum, \\ D--44780 Bochum, Germany} 
\\
$^{b}${\em Petersburg Nuclear Physics Institute, Gatchina, \\
St.Petersburg 188350, Russia} 
\\
$^{c}${\em RCNP, Osaka University, Mihogaoka 10-1, Ibaraki, \\
Osaka 567-0047, Japan}
\\
\end{center}
\vspace{1.5cm}
\begin{abstract}
\noindent
We summarize recent attempts to calculate the flavor asymmetry of the 
nucleon's sea quark distributions in the large--$N_c$ limit, 
where the nucleon can be described as a soliton of 
an effective chiral theory. We discuss the leading--twist longitudinally 
polarized and transversity antiquark distributions, 
$\Delta\bar u(x) - \Delta\bar d (x)$ 
and $\delta\bar u(x) - \delta\bar d (x)$, as well as the unpolarized
one, $\bar u(x) - \bar d (x)$, which appears only 
in the next--to--leading order of the $1/N_c$--expansion. 
Results for $\bar u(x) - \bar d (x)$ are in good agreement with the
recent Drell--Yan data from the FNAL E866 experiment. The 
longitudinally polarized antiquark asymmetry, 
$\Delta\bar u(x) - \Delta\bar d (x)$, 
is found to be larger than the unpolarized one.
\end{abstract}
\vfill
\rule{5cm}{.15mm} \\
{\footnotesize $\dagger$ Part of Plenary Talk presented
by C. Weiss at the XI International Conference ``Problems 
of Quantum Field Theory'' (In Memory of D.I.\ Blokhintsev),
Joint Institute for Nuclear Research, Dubna, Russia, July 13--17,
1998.}
\newpage
Hard scattering experiments have been a major tool for investigating
the structure of the nucleon, giving information about the distribution
of quarks, antiquarks and gluons in the nucleon. In the past decade the 
polarized parton distributions have been the main point of interest. 
Perhaps even more subtle are the flavor asymmetries of the quark, and, 
in particular, the antiquark distribution, the experimental study
of which has begun only recently.
\par
Flavor asymmetries can in principle be measured in deep--inelastic 
scattering. The difference
of proton and neutron structure functions measured by the NMC Collaboration 
has allowed to extract information
about the first moment of the flavor asymmetry of the unpolarized 
antiquark distribution in the proton \cite{NMC94},
\be
\int_0^1 dx \ [\bar{u}(x) - \bar{d}(x)] &=& -0.147 \pm 0.039 
\hspace{1cm} \mbox{at} \hspace{1cm} Q \;\; = \;\; 2\,{\rm GeV} ,
\label{NMC}
\ee
showing a large excess of $d$-- over $u$--antiquarks. 
This circumstance is frequently
expressed as a deviation of the so-called Gottfried sum,
\be
I_G &=& \frac{1}{3} + \frac{2}{3} \int_0^1 dx \ [\bar{u}(x) - \bar{d}(x)] ,
\label{Gottfried_sum}
\ee
from the value $1/3$ (Gottfried sum rule) \cite{Gottfried67,Kumano97}. 
Note that this sum rule does not follow from any fundamental principles 
of QCD. A more direct measurement of the antiquark distribution
is possible in Drell--Yan production. First limits on the 
flavor asymmetry of the unpolarized antiquark distribution were
obtained by the FNAL-711 experiment \cite{FNAL711}. The NA51 Collaboration 
at CERN has measured the ratio $\bar d / \bar u$ at a single value 
of $x$ \cite{NA51}. Recently, 
the E866 Experiment at FNAL has for the first time provided direct 
information about the $x$--dependence of the ratio $\bar d / \bar u$,
and thus about the shape of the unpolarized antiquark asymmetry,
over a wide range of $x$ \cite{E866,E866Peng}. 
\par
About the asymmetry of the
polarized antiquark distribution little is known at present from
experiment. In the parametrizations by Gl\"uck {\it et al.}\ 
of polarized structure function data 
$\Delta\bar{u}(x) - \Delta\bar{d}(x)$ was set to zero at the input 
scale \cite{GRSV96}. Also, flavor symmetry of the antiquark distribution
has been assumed in the extraction of the moments of the polarized
valence distribution from the SMC data for semi-inclusive spin
asymmetries \cite{SMC97}. One may hope that the spin structure experiments
planned at CERN (COMPASS \cite{COMPASS}), HERA and SLAC will provide
quantitative information about this asymmetry.
\par
It is clear that the large observed flavor asymmetry of the unpolarized 
sea quark distribution, Eq.(\ref{NMC}), cannot be explained by radiative 
generation of the antiquark distribution from some input valence
quark distribution at a low scale. Attempts of a theoretical explanation of 
the flavor asymmetry often appeal to the concept 
of a meson cloud of the nucleon, familiar from nuclear 
physics \cite{Kumano97,cloud}. For instance, a picture in which the
proton state has a component of a virtual $\pi^+$ and a ``core'' of
neutron quantum numbers, in which the virtual photon can scatter
off the pion (Sullivan mechanism \cite{Sullivan72}), could 
naturally explain the sign and overall magnitude
of the observed asymmetry \cite{Kumano97,cloud}. This
picture has been extended to include also contributions of the $\rho$
meson cloud of the nucleon to the asymmetry of the 
polarized antiquark distribution \cite{Fries98,BoreskovKaidalov98}. 
While intuitively appealing, it is difficult to maintain
a clear distinction between the contributions to the cross section
from the ``core'' and the ``cloud'' (see also Ref.\cite{KFS96} for a 
critical discussion).
\par
A more rigorous approach, which nevertheless retains the physical essence
of the ``meson cloud'' picture, is based on the large--$N_c$ limit
of QCD. It is well known that in the theoretical limit of a large
number of colors QCD becomes equivalent to an effective theory of mesons,
in which baryons appear as solitons, {\em i.e.}, classical solutions
characterized by a mean meson field \cite{Witten}. At low energies
the effective dynamics is described by the chiral Lagrangian for the pion, 
which appears as a Goldstone boson of the spontaneous breaking of 
chiral symmetry. The first realization of the idea of the nucleon as a 
soliton of the pion field by Skyrme \cite{Skyrme62} was 
using a particular choice of higher--derivative terms in the chiral
Lagrangian. A more realistic effective action, containing all orders 
in derivatives of the pion field, is defined by the integral over 
quark fields with a dynamically generated mass,
interacting with the pion field in a minimal chirally invariant way
\cite{DE}. Such an effective action has been derived from the instanton
vacuum of QCD, which provides a microscopic mechanism for the dynamical
breaking of chiral symmetry \cite{DP86}. It is valid in a wide range
of momenta up to the inverse instanton size, 
$\bar\rho^{-1} = 600\,{\rm MeV}$, which acts as an ultraviolet cutoff. 
The so--called chiral quark--soliton model of the nucleon based 
on this effective action \cite{DPP88} has been
very successful in describing hadronic observables such as the nucleon
mass, $N\Delta$--splitting, electromagnetic and weak form factors {\it
etc.} \cite{Review}. 
\par
The same approach allows to calculate also the leading--twist parton 
distributions of the nucleon at a low normalization 
point ($\mu \sim \bar\rho^{-1} = 600\, {\rm MeV}$)
\cite{DPPPW96,DPPPW97,PP96,PPGWW98}. The microscopic derivation of 
the effective chiral theory from the instanton model of the QCD vacuum
allows for a consistent identification of the twist--2 QCD
operators with operators in the effective theory \cite{DPW96}.
What is important is that the large--$N_c$ description of the nucleon as
a chiral soliton is fully field--theoretic and preserves 
all general properties of the parton
distributions, such as positivity and the partonic sum rules which hold in
QCD. In particular, it allows for a consistent calculation of the 
polarized and unpolarized antiquark
distributions, and thus of the flavor asymmetry. 
\par
The aim of this note is to give an overview of the results of 
Refs.\cite{DPPPW96,DPPPW97,PP96,PPGWW98}
for the flavor asymmetry of the antiquark distributions. 
We discuss the unpolarized as well as the longitudinally polarized 
and transversity antiquark distributions. For details we refer
to the original papers.
\par
The large $N_c$--limit implies a number of general statements
about the quark and antiquark distributions, which are independent
of the specifics of the low--energy dynamics. Quite generally, we aim
to describe parton distributions at values of $x$ parametrically
of order $x \sim 1/N_c$. On general grounds it can be shown that
the twist--2
distribution functions appearing in the leading order of the
$1/N_c$--expansion are the flavor--singlet unpolarized and the
flavor--nonsinglet longitudinally 
polarized one ($\Delta q$) \cite{DPPPW96,DPPPW97}, as well as
the flavor--nonsinglet transversity distribution ($\delta q$) \cite{PP96}.
They are of the form\footnote{For the definition of the polarized
distributions $\Delta u(x), \Delta d(x)$ implied here, 
see Ref.\cite{DPPPW96,DPPPW97}. We use $\delta u (x), \delta d (x)$
to denote $h_{1 u}(x), h_{1 d}(x)$ of Ref.\cite{PP96}. A general
discussion of transversity distributions can be found in 
Ref.\cite{Jaffe}.}
\be 
\left. 
\begin{array}{l} 
u(x) + d(x), \;\; \bar u(x) + \bar d(x) 
\\[.5cm] 
\Delta u(x) - \Delta d(x), \;\; \Delta \bar u(x) - \Delta \bar d(x)
\\[.5cm] 
\delta u(x) - \delta d(x), \;\; \delta \bar u(x) - \delta \bar d(x)
\end{array}
\right\}
&=&
N_c^2 \,\, F(N_c x),
\label{leading}
\ee
where $F(y)$ is a stable function in the large $N_c$--limit, which
depends on the particular distribution considered. The respective
other flavor combinations appear only in the next--to--leading
order of $1/N_c$ and are of the form
\be 
\left. 
\begin{array}{l} 
u(x) - d(x), \;\; \bar u(x) - \bar d(x) 
\\[.5cm] 
\Delta u(x) + \Delta d(x), \;\; \Delta \bar u(x) + \Delta \bar d(x)
\\[.5cm] 
\delta u(x) + \delta d(x), \;\; \delta \bar u(x) + \delta \bar d(x)
\end{array}
\right\}
&=&
N_c \,\, F(N_c x).
\label{subleading}
\ee
Thus, in the $1/N_c$--expansion the polarized flavor asymmetries are
parametrically larger than the unpolarized one. Note that this does not
necessarily imply that the polarized asymmetries are numerically larger;
for this one has to take into account the overall normalization of the 
distributions (see below).
\par
To actually calculate the (anti--) quark distributions at a low 
normalization point we need to
use the effective low--energy theory. Let us briefly sketch the
essential points of this approach (for details 
see Refs.\cite{DPPPW96,DPPPW97}).
In the effective chiral theory the nucleon is in the large $N_c$--limit
characterized by a classical pion field; in the nucleon rest frame it
is of ``hedgehog'' form,
\be
U ({\bf x}) &\equiv& e^{i\tau^a \pi^a ({\bf x})}
\;\;\; = \;\;\; e^{i\tau^a n^a P(r)} 
\label{hedge}
\ee
($n^a = x^a/|{\bf x}|, \; r = |{\bf x}|$),
where the profile function, $P(r)$, is determined by minimizing the
classical energy. Quarks are described by one--particle wave functions,
which are solutions of the Dirac equation in the background pion field,
\be
\gamma^0 \left( -i \gamma^k \partial_k + M e^{i\gamma_5\tau^a n^a P(r)} 
\right) \Phi_n ({\bf x}) &=& E_n \Phi_n ({\bf x}) .
\label{dirac}
\ee
Here, $M$ is the dynamical quark mass which arises in the spontaneous
breaking of chiral symmetry (numerically, 
$M \simeq 350\,{\rm MeV}$ \cite{DP86}). 
The spectrum of Eq.(\ref{dirac}) includes
a discrete bound--state level as well as a distorted negative and positive 
Dirac continuum. The discrete level and the negative continuum are occupied,
resulting in a state of unity baryon number. Nucleon states with
definite spin/isospin quantum numbers are obtained after quantizing the
rotational zero modes of the classical solution, Eq.(\ref{hedge}), which
are parametrized by
\be
U ({\bf x}) &\rightarrow& R(t) U ({\bf x}) R^\dagger (t) ,
\ee
with $R(t)$ an $SU(2)$ rotational matrix. An important point is that
the moment of inertia of the soliton is of order $N_c$, hence the
angular velocity is small, $\Omega = -iR^\dagger (dR/dt ) \sim 1/N_c$.
\par
The basic expressions for the quark and antiquark distributions
in this approach have been derived in Refs.\cite{DPPPW96,DPPPW97},
starting from the QCD definition of the distribution functions
as matrix elements
of certain light--ray operators in the nucleon, as well as from their
``parton model'' definition 
as the number of particles carrying a given fraction of the nucleon 
momentum in the infinite--momentum frame; both derivations lead to 
identical expressions for the distribution functions in the chiral
quark--soliton model. The $N_c$--leading distributions, Eq.(\ref{leading}), 
can be expressed
as sums of diagonal matrix elements of quark single--particle operators;
{\it e.g.}\ the flavor--nonsinglet polarized quark distribution
is given by
\be
\lefteqn{\Delta u (x) - \Delta d (x)} &&
\nonumber \\ 
&=& -\frac{1}{3} \, (2T_3) N_c M_N \,
\sum\limits_{\scriptstyle n\atop \scriptstyle{\rm occup.}}
\int\!\frac{d^3k}{(2\pi)^3}
\Phi_n^\dagger ({\bf k}) \; (1+\gamma^0\gamma^3) \, \gamma_5 \, \tau^3 \;
\delta(k^3 + E_n + xM_N) \; \Phi_n ({\bf k}) , \;\;\;\;\;\;
\label{isovector_occ}
\ee
where $\Phi_n ({\bf k})$ are the single particle wave functions,
Eq.(\ref{dirac}), in momentum representation, and $2T_3 = \pm 1$ 
for proton and neutron, respectively. Here the sum runs over all 
occupied quark single--particle levels ---
the bound state level and the negative continuum. The 
antiquark distribution is obtained as
\be
\Delta \bar u (x) - \Delta \bar d (x) 
&=& -\frac{1}{3}\, (2T_3) N_c M_N \,
\sum\limits_{\scriptstyle n\atop \scriptstyle{\rm occup.}} \;
\left\{ x \rightarrow -x \right\} .
\label{isovector_anti_occ}
\ee
Alternatively, it can be expressed as a sum over non-occupied levels 
({\it i.e.}\ the positive continuum),
\be
\Delta \bar u (x) - \Delta \bar d (x)
&=& \frac{1}{3} \, (2T_3) N_c M_N \,
\sum\limits_{\scriptstyle n\atop \scriptstyle{\rm non-occup.}} \;
\left\{ x \rightarrow -x \right\} .
\label{isovector_anti_nonocc}
\ee
The completeness of the set of quark single--particle wave 
functions is essential in ensuring correct properties of the quark
and antiquark distributions (sum rules {\it etc.}); 
see Refs.\cite{DPPPW96,DPPPW97} for a detailed discussion. For the
chirally--odd transverse polarized distributions the corresponding 
expressions are Eqs.(\ref{isovector_occ}), (\ref{isovector_anti_occ})
or (\ref{isovector_anti_nonocc})
with the matrix $\gamma_5$ replaced by $\gamma_5 \gamma^1 \tau_1$;
see Ref.\cite{PP96}.
\par
The flavor--nonsinglet unpolarized quark and antiquark distributions belong 
to the $N_c$--subleading ones, Eq.(\ref{subleading}). In the chiral
quark--soliton model this manifests itself in the fact that the
expressions for the nucleon matrix elements of the light--cone operator
become non-zero only  
after expanding to first order in the angular velocity of the soliton; 
as a result the distributions are given by double
sums over quark single--particle levels, similar to the moment
of inertia of the classical soliton. We do not quote the
lengthy expressions here but rather refer to Ref.\cite{PPGWW98}. 
\par
Eqs.(\ref{isovector_occ}), (\ref{isovector_anti_occ}) or 
(\ref{isovector_anti_nonocc}) serve as a starting point for a numerical
evaluation of the distribution functions. A straightforward way to
compute the distributions is to diagonalize the
hamiltonian, Eq.(\ref{dirac}), and perform the
sum over contributions of single--particle levels 
numerically \cite{DPPPW97}.
Also, it is worthwhile to note that an (almost) analytic answer for the 
distribution functions can be obtained in the hypothetical limit of large 
soliton size, which allows one to perform an expansion in the inverse 
soliton size, analogous to the usual ``gradient expansion'' for nucleon matrix 
elements of local operators; see Refs.\cite{DPPPW96,DPPPW97} for
details.
\par
\begin{figure}
\setlength{\epsfxsize}{15cm}
\setlength{\epsfysize}{15cm}
\epsffile{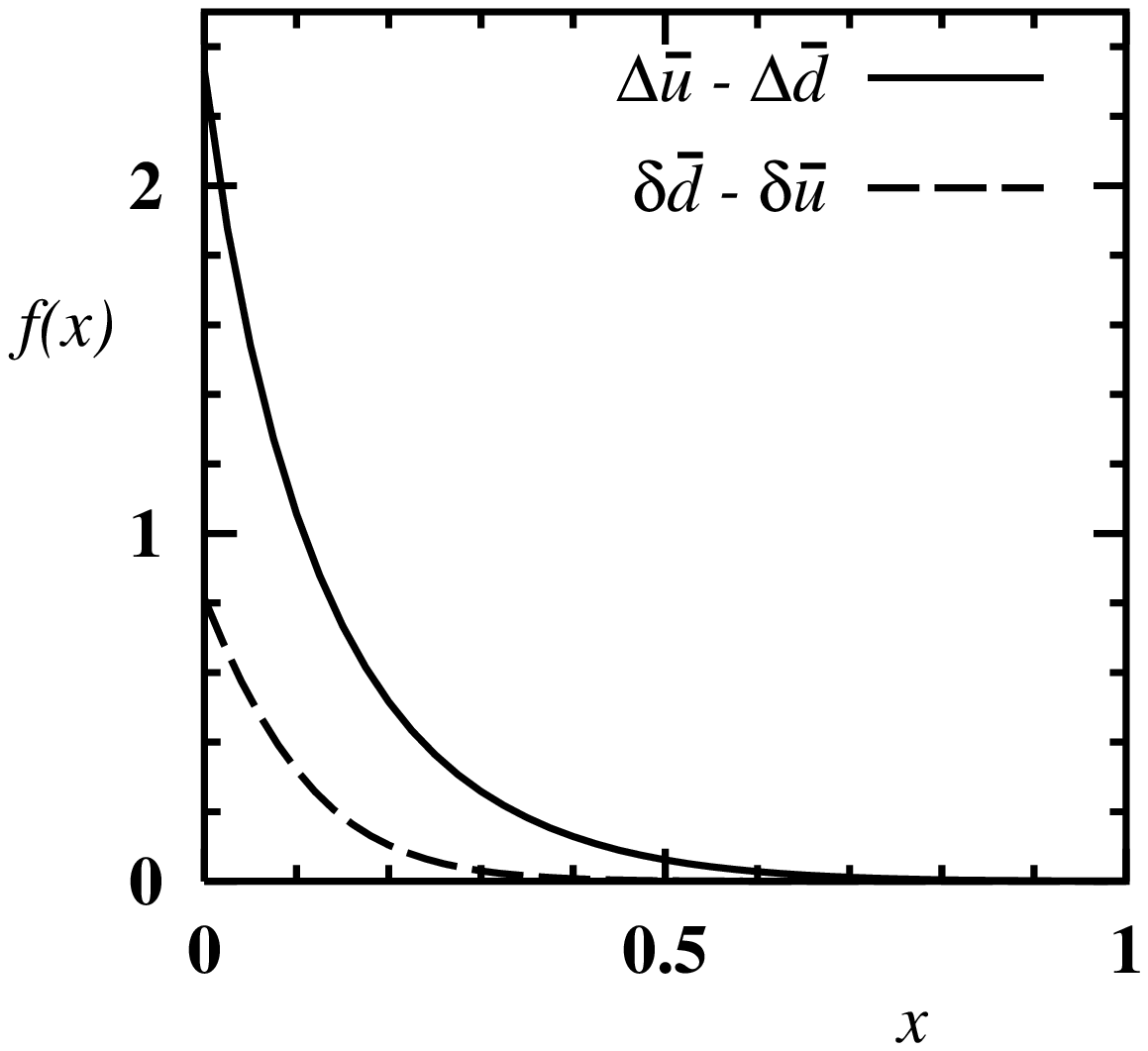}
\caption[]
{The calculated flavor asymmetry of the longitudinally polarized antiquark 
distribution in the proton, $\Delta\bar u(x) - \Delta\bar d (x)$ 
({\it solid line}), and the transversity antiquark distribution, 
$\delta\bar d(x) - \delta\bar u(x)$ ({\it dashed line}), at the
low normalization point.}
\label{fig_aspol}
\end{figure}
\begin{figure}
\setlength{\epsfxsize}{15cm}
\setlength{\epsfysize}{15cm}
\epsffile{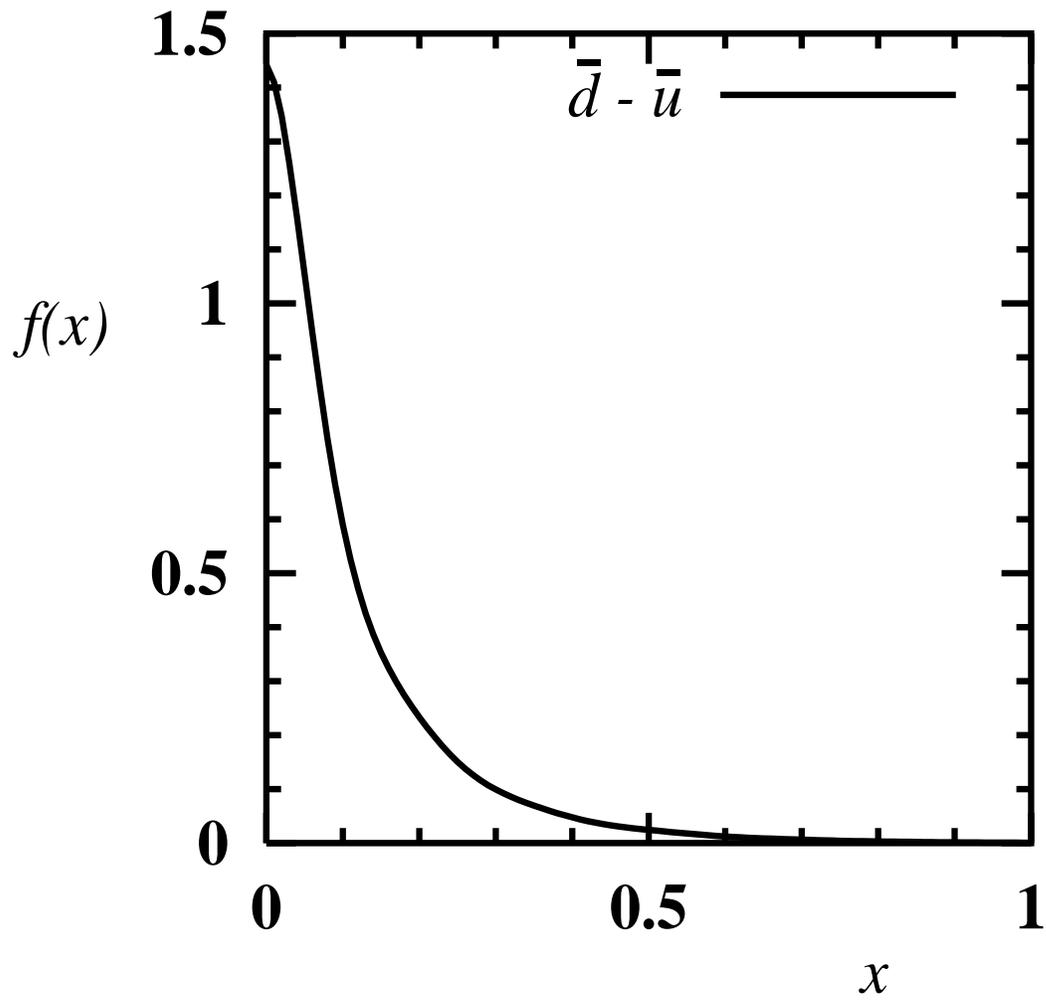}
\caption[]
{The calculated unpolarized antiquark asymmetry in the proton, 
$\bar d(x) - \bar u(x)$, at the low normalization 
point \cite{PPGWW98}.}
\label{fig_asnon}
\end{figure}

The results for the flavor asymmetries of the antiquark distributions
are shown in 
Figs.\ref{fig_aspol} and \ref{fig_asnon}. The polarized antiquark 
asymmetries at the low normalization point 
($\mu \sim \bar\rho^{-1} = 600\,{\rm MeV}$), which are
leading in the $1/N_c$ expansion, are shown in 
Fig.\ref{fig_aspol}. As can be seen, both the longitudinally 
polarized as well as the transversity asymmetry have definite 
sign\footnote{In Fig.\ref{fig_aspol} we quote results obtained
with the variational (arctan--) soliton profile of 
Refs.\cite{DPPPW96,DPPPW97},
using the ``interpolation formula'' and a Pauli--Villars ultraviolet
cutoff applied to the Dirac continuum contribution. The contribution
of the discrete level is not regularized.}. The sign
of our result is in agreement
with the $\rho$ meson cloud model of Ref.\cite{Fries98}
(note that these authors are using a definition of the polarized
antiquark distribution with sign opposite to ours); 
however, the polarized asymmetry obtained in our approach is larger 
by almost an order of magnitude. It would be extremely interesting
to incorporate this asymmetry in analyses of experimental data,
{\em e.g.}\ the SMC data for semi-inclusive spin asymmetries \cite{SMC97}.
For the first moments of the flavor--nonsinglet antiquark distributions
we obtain
\be
\int_0^1 dx \left[ \Delta\bar u (x) - \Delta\bar d (x) \right]
&=& 0.31,
\\
\int_0^1 dx \left[ \delta\bar u (x) - \delta\bar d (x) \right]
&=& -0.082 .
\ee
As already said, these values should be associated with a normalization
point of the order $\mu \sim \bar\rho^{-1} = 600\, {\rm MeV}$. 
\par
We now turn to the unpolarized antiquark asymmetry. The result of the
model calculation for $\bar u(x) - \bar d(x)$ at the low normalization
point \cite{PPGWW98} is shown in Fig.\ref{fig_asnon}.\footnote{A 
calculation of this distribution in a related approach has been 
reported in Ref.\cite{WK97}; see Ref.\cite{PPGWW98} for a discussion
of differences.} The first moment of the calculated distribution 
at $\mu \sim \bar\rho^{-1} = 600\, {\rm MeV}$ is
\be
\int_0^1 dx \left[ \bar u (x) - \bar d (x) \right]
&=& -0.17 .
\ee
Since this quantity exhibits only very weak scale dependence
it is justified to compare this directly with
the NMC value at $Q = 2\,{\rm GeV}$, Eq.(\ref{NMC}). We see that
our value is consistent with the NMC result. We note that the E866 
Drell--Yan data for $\bar d(x) / \bar u (x)$ \cite{E866}, combined with the 
CTEQ4M parametrization of $\bar u(x) + \bar d(x)$ \cite{CTEQ97}, 
suggest a value
for the integral Eq.(\ref{NMC}) of about 2/3 the NMC result, which, however,
depends on the parametrization of the parton distributions used
to estimate the contributions from the unmeasured region $x < 0.02$; 
see Ref.\cite{E866Peng} for a detailed discussion of the compatibility of 
this result with the NMC measurement. Note also that there are 
systematic uncertainties in our model calculation related to the use of 
the $1/N_c$--expansion as well as the lack of knowledge of the precise 
form of the ultraviolet cutoff
of the effective chiral theory \cite{DPPPW96,DPPPW97}.
\par
The first moment of the flavor asymmetry of the unpolarized antiquark 
distribution (the Gottfried sum) has been studied previously in the Skyrme 
model \cite{WH93} and the chiral quark--soliton model \cite{Wakamatsu92}.
These calculations attempted to calculate the Gottfried sum directly,
using certain operator expressions for this quantity which were 
not derived from a consistent identification of the parton distribution 
functions in the low--energy model.
\par
\begin{figure}
\setlength{\epsfxsize}{15cm}
\setlength{\epsfysize}{15cm}
\epsffile{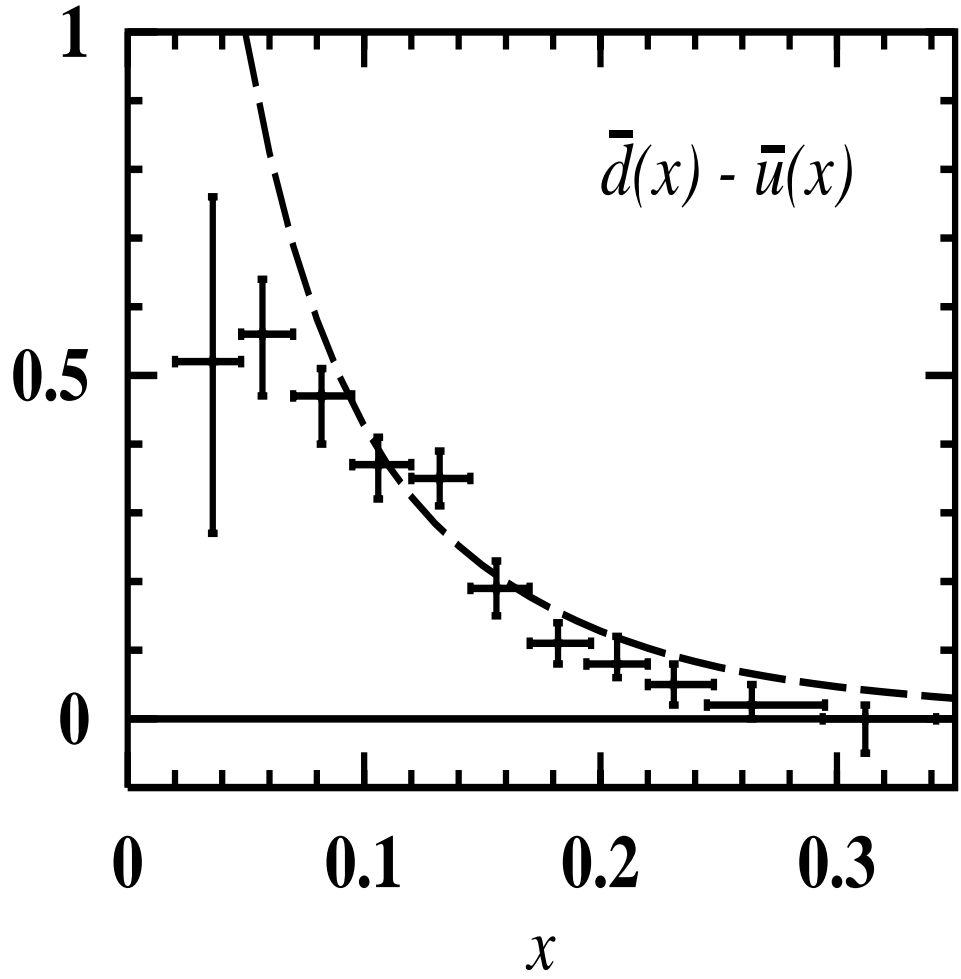}
\caption[]
{The values for $\bar d(x) - \bar u(x)$ in the proton at $Q = 7.35\,{\rm GeV}$
from the analysis of the FNAL E866 data of Ref.\cite{E866Peng}, 
compared to the distribution calculated in 
Ref.\cite{PPGWW98}, evolved from $\mu = 600\,{\rm MeV}$ to the 
experimental scale ({\it dashed line}).}
\label{fig_e866}
\end{figure}

For the $x$--dependence of the unpolarized antiquark asymmetry data 
are available from 
the Fermilab E866 Drell--Yan experiment \cite{E866}. We cannot directly 
compare the measured ratio $\bar d(x) / \bar u(x)$ to the model calculation,
since this quantity is inhomogeneous in the parameter $1/N_c$, and to 
compute it we would need to know the flavor--singlet distribution, 
$\bar u(x) + \bar d(x)$, in next--to--leading order of the 
$1/N_c$--expansion, {\it cf.}\
Eqs.(\ref{leading}) and (\ref{subleading}). We therefore compare 
$\bar u (x) - \bar d (x)$, which was extracted from the E866 data
for $\bar d(x) / \bar u(x)$
combined with the CTEQ4M parametrization of $\bar u(x) + \bar d(x)$.
Fig.\ref{fig_e866} shows the data for $\bar d (x) - \bar u (x)$ extracted 
from the analysis of Ref.\cite{E866Peng}, together
with the result of the calculation in the chiral quark--soliton 
model of Ref.\cite{PPGWW98}. Here we have evolved the distribution
calculated in Ref.\cite{PPGWW98} from the low normalization point
($\mu \sim 600\, {\rm MeV}$) to the scale of $Q = 7.35\,{\rm GeV}$,
using leading--order evolution with $\Lambda_{\rm QCD} = 232\,{\rm MeV}$
for $N_f = 3$. We remark that the results of the present model are not 
meaningful for 
small $x$, since for values of $x$ parametrically of the order 
$(M\bar\rho )^2/ N_c$ ($\bar\rho^{-1}$ is the inverse average
instanton size, {\it cf.}\ above) effects not taken into account in the 
present calculation become important; see \cite{PPPBGW97,PPGWW98} for 
details.
\par
Of interest is also the comparison of the integral
of this distribution over the measured $x$--region, $0.02 < x < 0.345$, with
the result of the model calculation. After evolution of the calculated
antiquark distribution we find
\be
\int_{0.02}^{0.345} dx \left[ \bar u (x) - \bar d (x) \right]
&=& -0.108 \hspace{1cm} \mbox{at} \hspace{1cm} Q \;\; = \;\; 7.35\,{\rm GeV},
\ee
to be compared with the value 
$-0.068 \pm 0.007 (\mbox{stat.}) \pm 0.008 (\mbox{syst.})$ obtained
in the analysis of Ref.\cite{E866Peng} (see that paper for 
details). For the first moment we obtain
\be
\int_{0.02}^{0.345} dx \,\, x \,\left[ \bar u (x) - \bar d (x) \right]
&=& -0.0096 \hspace{1cm} \mbox{at} \hspace{1cm} Q \;\; = \;\; 7.35\,{\rm GeV},
\ee
to be compared with $-0.0065 \pm 0.0010$ \cite{E866Peng}.
\par
To summarize, we have shown that the large--$N_c$ picture of the
nucleon as a chiral soliton naturally gives a flavor asymmetry 
of the unpolarized antiquark distribution in agreement with 
the observed violation of the Gottfried sum rule, and with the 
recent first results for the $x$--dependence of the asymmetry from 
Drell--Yan production. Equally important, this picture
predicts a sizable asymmetry of the polarized antiquark distribution
(both longitudinally polarized and transversity distribution).
It would be extremely interesting to incorporate this information
in new parametrizations of the parton distribution functions,
or directly in the analyses of experimental data.
%

%
%
\newpage
\newpage
\newpage
\end{document}